\documentstyle[preprint,prd,floats,eqsecnum,aps,epsf]{revtex}

\begin{document}
\preprint{\vbox{\hbox {October 1998} \hbox{IFP-763-UNC} } }
\title{\bf Aspects of Soft and Spontaneous CP Violation.}
\author{\bf Paul H. Frampton and Masayasu Harada}
\address{Department of Physics and Astronomy,}
\address{University of North Carolina, Chapel Hill, NC  27599-3255}
\maketitle

\begin{abstract}
We study four different models for CP violation: the standard (KM) 
model, the aspon model of 
spontaneous breaking and two models of soft breaking. In all 
except the standard model,
the strong CP problem is addressed and solved. Testable predictions
for the area of the unitarity triangle and for 
$(\epsilon^{'}/\epsilon)_K$
are emphasized. 
The issue of CP violation may well become the first place
where the standard model of particle theory is shown definitively
to be deficient. There are two reasons for expecting this to happen:
(1) the strong CP problem is still not 
understood in the unadorned standard model
and (2) the KM mechanism, although unquestionably present, may not
provide the full explanation of $\epsilon_K$ and 
$(\epsilon^{'}/\epsilon)_K$.
\end{abstract}


\newpage

\section{Introduction}

There are several models for describing CP violation by scalar
dynamics.  Spontaneous CP violation~\cite{W,B} is one of the
interesting schemes, especially where CP is broken simultaneously with
SU(2)$_L\times$U(1)$_Y$.  This kind of models has been widely 
studied (see {\it e.g.} Refs.~\cite{S,BS,CGN}).

There are other models in which heavy quarks and scalars are
introduced and CP violation is originated in the heavy scalar sector.
The CP violation is transported to the ordinary quark sector through
the Yukawa interactions among heavy quark, ordinary quark and heavy
scalar. At the same time, an attempt is made to resolve the Strong CP
problem.
These models may be divided into two classes by the existence of the
tree level flavor changing neutral current (FCNC).

There are two typical models without tree level FCNC.
In one class of models only right-handed quarks have Yukawa
interactions with heavy quarks and scalars~\cite{BCK}, while in
another class only left-handed quarks have the Yukawa
interaction~\cite{GG}.  In both models CP is violated in the heavy
mass terms softly or spontaneously.

A typical model with the tree level FCNC is the Aspon
model~\cite{Aspon}.
This model is widely studied (see, {\it e.g.} Refs.~\cite{FN,AFKL}).
In this model one vector-like SU(2)$_L$ doublet of quarks are
introduced.
Those quarks have same charges as up- and down-types of ordinary
quarks.  Two heavy SU(2)$_L\times$U(1)$_Y$ singlet scalars have VEVs
which break CP spontaneously.
Another model with tree level FCNC is given in \cite{BBP},
where unlike the models considered here, no additional U(1) symmetry
occurs.

In this paper we study constraints and predictions of the above two
models of soft CP breaking comparing with those for the Aspon model.
These considerations are timely because experiments are underway
to measure both Re$(\epsilon'/\epsilon)$ and the CP asymmetries
in $B^0$ decays. In fact, there are two experiments each to measure
both effects. The former is being measured by the NA-48 experiment
at CERN, and by the E799/E832 (KTEV) experiment at FermiLab.  The
latter is being studied by the BaBar detector of the PEP-II experiment 
at SLAC and by the BELLE experiment at the KEK B-Factory.

The layout of the paper is as follows: In Section(\ref{Models})
we describe in detail the different models we shall analyze.
In Section(\ref{epsilon}) the constraints arising from
$\epsilon_K$ are derived.  The predictions for
$\epsilon^{\prime}/\epsilon$ are given in Section(\ref{prime}).
In Section(\ref{B}) the constraint of $B-\bar{B}$ mixing
is discussed. Section(\ref{decays}) covers the CP asymmetry 
predictions for neutral $B$ decay. In Section(\ref{theta}) the
lower limits on $\bar{\Theta}$ are calculated, together with the 
corresponding lower limits on the neutron electric dipole moment.
Finally in Section(\ref{summary}) the different predictions
are summarized.

\section{Models}   \label{Models}

Here we shall list four different models for CP violation
which exemplify all of the ideas we are considering.
At the end of the paper we shall summarize the similarities and
differences
of the experimental predictions. Thus the hurried reader could read
just this section and that summary to sample the main points: the
intervening sections provide technical details.

\subsection{Standard Model}

The first model is just the standard model (SM)
with the KM mechanism~\cite{KM}
of explicit CP violation. Principally, we are interested in
models which also solve strong CP (as all the other three will). The
standard model requires an additional mechanism ({\it e.g.} the
Peccei-Quinn mechanism~\cite{PQ} or massless up quark (see, {\it e.g.}
Ref.~\cite{P}) to accomplish this. Nevertheless, it 
forms an essential comparison for all the other cases.

\subsection{Two Models (Types L and R) of Soft CP Breaking}

The class of models we consider for soft 
CP violation is constructed by adding two SU(2)$_L$ singlet
scalars $\chi_I$ ($I=1$, $2$) with hyper-charge $Y_\chi$
and one non-chiral quark $Q$ to the standard model (SM).
These $Q$ and $\chi_I$ carry the opposite charges of 
an extra U(1)$_{\rm new}$ symmetry.
The hypercharge of $Q$ is determined in such a way that the Yukawa
interactions among $\chi_I$, $Q$ and the ordinary quark $q$
are allowed: $Y_Q = Y_\chi + Y_q$.
The Yukawa interactions in the models can be written as
\begin{equation}
{\cal L}_Y = - \sum_{I=1}^2 \sum_{i=1}^3 h^I_i
\left[ 
  \overline{Q} q_{i} \chi_I  + \overline{q}_{i} Q \chi^\ast_I
\right]
\ ,
\end{equation}
where $h^I_i$ is a real Yukawa coupling.

CP is softly broken by the mass term of
$\chi_I$.
The models in this category are divided into two types by
the chirality of the ordinary quark $q$ which couples to $Q$ and
$\chi_I$:
in the first type (Type R), $q$ is a right-handed down-type quark, 
$Y_q = -1/3$~\cite{BCK};
in the second type (Type L), $q$ is a left-handed SU(2)$_L$ doublet
quark, $Y_q = 1/6$~\cite{GG}.

Let us explain details of these models for soft CP violation.
The scalar potential for $\chi_I$ is given by
\begin{equation}
{\cal L}_{\chi} =
\sum_{I,J,K,L=1}^2 \overline{\lambda}_{IJKL}
\chi_I^\ast \chi_J \chi_K^\ast
\chi_L + \sum_{I,J=1}^2 \overline{M}_{IJ} \chi_I^\ast \chi_J
\ ,
\label{L chi}
\end{equation}
where $\overline{\lambda}_{IJKL}$ and 
$\overline{M}_{II} = \overline{M}_{II}^{*}$ are real quantities
and $\overline{M}_{12}=\overline{M}^\ast_{21}$ is a
complex quantity.
The interaction between $\chi_I$ and the ordinary 
SU(2)$_L$ doublet Higgs scalar $H$ is given by
\begin{equation}
{\cal L}_{\chi H} = 
\left( H^{\dag} H -\frac{ v^2}{2} \right) \, 
\sum_{I,J=1}^2 \lambda_{IJ}
\chi^{\ast}_I \chi_J
\ ,
\label{L chi H}
\end{equation}
where $\lambda_{IJ}$ is real.
The mass eigenstate $\chi'_I$ is given by a unitary rotation:
\begin{equation}
\chi'_I = \sum_{J=1}^2 U_{IJ} \chi_J \ ,
\label{rotation}
\end{equation}
where $U$ is a suitable unitary matrix.

After rotating $\chi$ to the mass eigenstate $\chi'$ as in
Eq.~(\ref{rotation}), the Yukawa interactions become
\begin{equation}
{\cal L}_Y = - \sum_{I=1}^2 \sum_{i=1}^3 
\left[ 
  f_{Ii} \overline{Q} q_i \chi'_I 
  + f_{Ii}^\ast \overline{q}_i Q \chi^{\prime\ast}_I
\right]
\ ,
\end{equation}
where $f_{Ii} = \sum_{J=1}^2 U^\ast_{IJ} h^J_i$ is a complex Yukawa
coupling.

An important combination for CP measurement is 
$X^I_{ij} \equiv f^\ast_{Ii}f_{Ij}$.
The fact that the original Yukawa coupling $h^I_i$ is real leads
\begin{equation}
\mbox{Im} \left( X^{I=2}_{ij} \right)
= - \mbox{Im} \left( X^{I=1}_{ij} \right)
\ .
\label{cond:Imf}
\end{equation}

\subsection{Aspon Model}  

Here the complex scalar $\chi_I$ has a Vacuum Expectation Value (VEV)
which spontaneously breaks CP (Aspon model~\cite{Aspon});
In the Aspon model the U(1)$_{\rm new}$ is gauged,
and the gauge boson acquires its mass from the VEV of $\chi_I$.
$Q$ and $q$ have same charges ($Y_\chi=0$, $Y_Q=Y_q$),
and they have
complex mass mixing.  Accordingly, there exist tree-level flavor
changing neutral currents (FCNC) mediated by the Aspon gauge boson and
$\chi_I$.

Let us briefly review the relevant part of the 
Aspon model.
In the Aspon model $q$ can be the left-handed doublet
quarks, or the right-handed down-type quarks, in the simplest versions.
In the present analysis we fix $q$ to be the left-handed
doublet quarks for definiteness.\footnote{%
In the concluding section VIII we mention the difference
in predictions for an R-type Aspon model.}
All the couplings in the scalar potentials in 
Eqs.~(\ref{L chi}) and (\ref{L chi H}) are real, and CP
is spontaneously broken by the VEV of $\chi_I$:
\begin{equation}
\left\langle \chi_1 \right\rangle = \frac{1}{\sqrt{2}}
\kappa_1 e^{i\theta} \ , \quad
\left\langle \chi_2 \right\rangle = \frac{1}{\sqrt{2}}
\kappa_2 \ .
\end{equation}
As a result the light quarks $q$ mix with the non-chiral
heavy quark $Q$.  The mass matrix, in the weak basis where
$3\times3$ submatrix for down sector is diagonal, is given
by~\cite{FN}
\begin{equation}
{\cal M}_d = 
\left( \begin{array}{cccc}
m_d & 0 & 0 & F_1 \\
0 & m_s & 0 & F_2 \\
0 & 0 & m_b & F_3 \\
0 & 0 & 0 & M_Q
\end{array} \right) \ ,
\end{equation}
where
\begin{equation}
F_i = h^1_i \left\langle \chi_1 \right\rangle
+ h^2_i \left\langle \chi_2 \right\rangle \ .
\end{equation}
This mass matrix is diagonalized by a biunitary
transformation, $K_L^{\dag} {\cal M}_d K_R$.
The approximate form of the transformation matrices are given 
by~\cite{FN}
\begin{eqnarray}
K_L &=&
\left( \begin{array}{cccc}
1 - \frac{1}{2} \left\vert x_1 \right\vert^2 &
 x_1 x_2^\ast \frac{m_s^2}{m_d^2-m_s^2} &
 x_1 x_3^\ast \frac{m_b^2}{m_d^2-m_b^2} &
 x_1
\\
x_2 x_1^\ast \frac{m_d^2}{m_s^2-m_d^2} &
 1 - \frac{1}{2} \left\vert x_2 \right\vert^2 &
 x_2 x_3^\ast \frac{m_b^2}{m_s^2-m_b^2} &
 x_2
\\
x_3 x_1^\ast \frac{m_d^2}{m_b^2-m_d^2} &
 x_3 x_2^\ast \frac{m_s^2}{m_b^2-m_s^2} &
 1 - \frac{1}{2} \left\vert x_3 \right\vert^2 &
 x_3
\\
 - x_1^\ast &  - x_2^\ast & - x_2^\ast &
 1 - \frac{1}{2} \sum_{j=1}^3 \left\vert x_j \right\vert^2
\end{array}
\right)
\ ,
\nonumber\\
K_R &=&
\left( \begin{array}{cccc}
1 &
 x_1 x_2^\ast \frac{m_d m_s}{m_d^2-m_s^2} &
 x_1 x_3^\ast \frac{m_d m_b}{m_d^2-m_b^2} &
 \frac{m_d}{M_Q} x_1
\\
x_2 x_1^\ast \frac{m_s m_d}{m_s^2-m_d^2} &
 1 &
 x_2 x_3^\ast \frac{m_s m_b}{m_s^2-m_b^2} &
 \frac{m_s}{M_Q} x_2
\\
x_3 x_1^\ast \frac{m_b m_d}{m_b^2-m_d^2} &
 x_3 x_2^\ast \frac{m_b m_s}{m_b^2-m_s^2} &
 1 &
 \frac{m_b}{M_Q} x_3
\\
 - \frac{m_d}{M_Q} x_1^\ast &
 - \frac{m_s}{M_Q} x_2^\ast &
 - \frac{m_b}{M_Q} x_3^\ast &
 1 
\end{array}
\right)
\ ,
\end{eqnarray}
where 
\begin{equation}
x_i \equiv F_i/M_Q \ .
\label{def x}
\end{equation}

In the weak basis the Aspon gauge boson does not
couple to light quarks.  However, due to the mixing
with the heavy quark $Q$, light quarks in terms of the mass
eigenstates couple to the Aspon gauge boson.
This induces FCNC's:
\begin{equation}
{\cal L}_A^{\rm FCNC} (\mbox{down}) = - g_A
\alpha_{ij} \bar{d}^{\prime i}_L \gamma_\mu d^{\prime j}_L
A^\mu \ ,
\end{equation}
where
\begin{eqnarray}
&& \alpha_{ij} \simeq x_i x_j^\ast \ ,
\qquad \mbox{($i,j= 1,2,3$)} \ ,
\nonumber\\
&& \alpha_{4i} = \alpha_{i4}^\ast \simeq - x_i^\ast \ ,
\qquad \mbox{($i= 1,2,3$)} \ ,
\nonumber\\
&& \alpha_{44} \simeq 1 - \sum_{i=1}^{3} 
\left\vert x_i \right\vert^2
\ ,
\end{eqnarray}
with $A^\mu$ being the Aspon gauge boson and $g_A$
the gauge coupling.  
In addition to the above FCNC's in the left-handed sector
there exist FCNC's in the 
right-handed sector.
However, the coupling is suppressed by the mass ratio
$m_i/M_Q$, where $m_i=(m_d,\,m_s,\,m_b)$.
Similarly, flavor changing couplings to $\chi_I$ are suppressed 
by $m_i/M_Q$.
So we will neglect these couplings below.

\section{Constraint from $\epsilon_K$}
\label{epsilon}

In the SM, $\epsilon_K$ arises from the $W^+W^-$ exchange
box diagram, and is proportional to a combination of CKM
angles and to $\sin\delta$ where $\delta$ is the KM phase, and
therefore gives a constraint between these SM parameters.

Now we study the other models defined in Section(\ref{Models}).
The parameter $\epsilon_K$ is given by
\begin{equation}
\epsilon_K = 
\frac{e^{i\pi/4}}{2\sqrt{2}}
\left[ \frac{\mbox{Im}M_{12}}{\mbox{Re}M_{12}} + 2
  \frac{\mbox{Im}A_0}{\mbox{Re}A_0}
\right]
\ .
\end{equation}
The second term is related to $\epsilon'/\epsilon$, and much smaller
than the first term as we shall see below.

\begin{figure}[thbp]
\begin{center}
\ \epsfbox{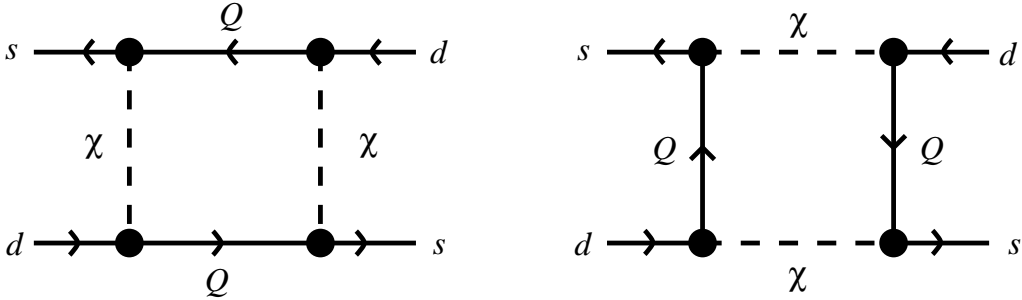}
\end{center}
\caption[]{Box diagram contributions to $K^0$--$\bar{K}^0$ mixing in
the models of soft CP breaking.
}\label{fig:box}
\end{figure}
\begin{figure}[thbp]
\begin{center}
\ \epsfbox{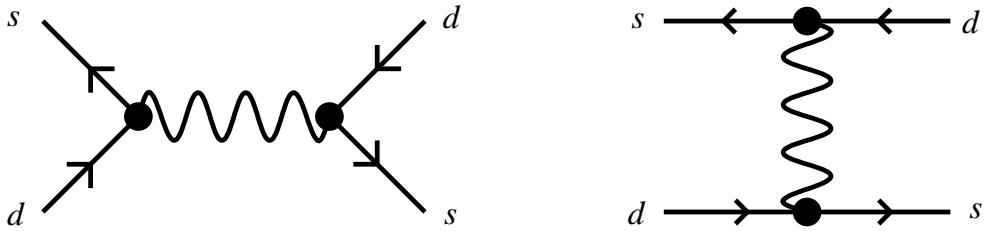}
\end{center}
\caption[]{Tree level Aspon gauge boson exchange contributions to 
$K^0$--$\bar{K}^0$ mixing in the Aspon model.
}\label{fig:aspon}
\end{figure}

The dominant contribution to $\mbox{Im}M_{12}$ is given by the
scalar-heavy quark exchange box diagram shown in Fig.~\ref{fig:box}
for the models of soft CP
breaking, and by the Aspon gauge boson exchange tree diagram 
shown in Fig.~\ref{fig:aspon} for the
Aspon model.
The effective $\Delta S = 2$ Hamiltonian derived from the contribution,
for Type R soft breaking, is given by
\begin{equation}
{\cal H}_{\Delta S =2}^{\rm (new)}
= \frac{1}{v^2} C_{sd}^{(R)} 
\left( \bar{s}_R \gamma^\mu d_R \right)
\left( \bar{s}_R \gamma_\mu d_R \right) \ ,
\label{new H}
\end{equation}
where
\begin{equation}
C_{sd}^{(R)} = \frac{1}{6(4\pi)^2}
\frac{v^2}{M_Q^2} \sum_{I,J=1}^{2} X^I_{sd} X^J_{sd}
\, F\left( r_I , r_J  \right) \ , \label{Csd}
\end{equation}
with $r_I = M_I^2/M_Q^2$.
The function $F(r_I,r_J)$ is defined by
\begin{equation}
F\left( r_I, r_J \right) =
\frac{3}{ (1-r_I) (1-r_J)} - 
\frac{3r_J^2}{(1-r_J)^2 (r_I-r_J)} \ln r_J +
\frac{3r_I^2}{(1-r_I)^2 (r_I-r_J)} \ln r_I
\ ,
\end{equation}
where the normalization of $F(r_I,r_J)$ is taken as
$F(1,1)=1$.
For Type L soft breaking, and for the Aspon model, the effective
coupling is the same as Eq.~(\ref{new H}) with the helicities switched
from R to L, and with the coefficient $C_{sd}^{(R)}$ 
replaced by $C_{sd}^{(L)}$, and 
$C_{sd}^{(A)}$, respectively. The formula for $C_{sd}^{(L)}$
is exactly as for $C_{sd}^{(R)}$ in Eq.(\ref{Csd}). 
We will give the formula for $C_{sd}^{(A)}$ later.

As in the SM, $\mbox{Re}M_{12}$ is dominated by the contribution 
from $W$-charm exchange box diagram.
This is given by
\begin{equation}
{\cal H}_{\Delta S =2}^{\rm (KM)}
= \frac{1}{v^2} C_{sd}^{\rm (KM)}
\left( \bar{s}_L \gamma^\mu d_L \right)
\left( \bar{s}_L \gamma_\mu d_L \right) \ ,
\label{KM H}
\end{equation}
where
\begin{equation}
C_{sd}^{\rm (KM)} = \frac{1}{8\pi^2}
\frac{M_W^2}{v^2}
\left( V_{cs}^\ast\, V_{cd} \right)^2
S\left( \frac{m_c^2}{M_W^2} \right) \ .
\end{equation}
The function $S(x)$ is so-called Inami-Lim function~\cite{IL} and
$S(m_c^2/M_W^2) \simeq 3.48 \times 10^{-4}$.
$V_{cs}$ and $V_{cd}$ are corresponding elements of the quark mixing 
matrix, $\left\vert V_{cs}^\ast V_{cd} \right\vert \simeq 0.22 $.
Note that the mixing matrix for the models of soft CP
breaking is real and orthogonal, and the $3 \times 3$ submatrix
of it in the Aspon model is also real and orthogonal to a
very good approximation\cite{FG}.

Since QCD respects parity invariance, it may be enough to assume that
two operators in Eqs.~(\ref{new H}) and (\ref{KM H}) give the same
hadron matrix elements.
Then $\left\vert \epsilon_K \right\vert$ can be expressed as
\begin{equation}
\left\vert \epsilon_K \right\vert \simeq
\frac{1}{2\sqrt{2}} 
\left\vert \frac{\mbox{Im}C_{sd}^{(R,L,A)}}{C_{sd}^{\rm (KM)}}
\right\vert \ .
\end{equation}
The experimental value $\left\vert \epsilon_K \right\vert = 2.26
\times 10^{-3}$ gives a constraint to $\left\vert \mbox{Im}C_{sd}
\right\vert$:
\begin{equation}
\left\vert \mbox{Im}C_{sd}^{(R,L,A)}\right\vert \simeq
1.4 \times 10^{-10} \ .  \label{eps}
\end{equation}
This smallness of $\left\vert \mbox{Im}C_{sd}^{(R,L,A)} \right\vert$
is easily understood by small Yukawa coupling $f_{Ii}$.

To estimate the size of the Yukawa couplings, we can assume that their
real and imaginary parts are comparable, equate $M_Q$ and $M_I$ 
and arrive, from Eq.(\ref{Csd}), at:
\begin{equation}
\frac{v}{M_Q} \bar{X}_{sd}^{(R,L)} \simeq 3 \times 10^{-4} \ ,
\label{X}
\end{equation}
where we have defined $(\bar{X}_{sd})^{2} = \frac{1}{2}
|\mbox{Im}X_{sd}^{I=1}
\sum_{I=1}^{2}
\mbox{Re}X_{sd}^{I}|$ using the fact that $\mbox{Im}X_{sd}^{I=2} = 
- \mbox{Im}X_{sd}^{I=1}$.
Of course, the corresponding Yukawa couplings involving the third
family, {\it e.g.} $\bar{X}_{bd}$, $\bar{X}_{bs}$ are not constrained
by $\epsilon_K$.
It seems natural to say that $M_Q$ is bigger than the weak scale, and
then Eq.~(\ref{X}) gives the lower bound 
$\bar{X}_{sd}^{(R,L)} \gtrsim 3 \times 10^{-4}$.

In the Aspon model
\begin{equation}
C_{ds}^{(A)} = 
2 \left( \frac{v}{\kappa} \right)^{2} (x_1^*x_2)^2 \label{CA}
\end{equation} 
where $\kappa$ is the scale of U(1)$_{\rm new}$ breaking. The
combination of Eq.~(\ref{eps}) and Eq.~(\ref{CA}), as is 
well-known~\cite{FN,FH}, gives a
constraint on $\kappa$, using information from $\bar{\Theta}$
(see Section(\ref{theta})).
The parameter $x_3$ is not constrained by $\epsilon_K$.

\section{Predictions for Real Part of $(\epsilon^\prime/\epsilon)$}
\label{prime}

In the standard model, an enormous effort has gone into
calculating direct CP violation, characterized by the quantity
$\mbox{Re}\left( \epsilon' / \epsilon \right)$
(see, {\it e.g.} Refs.~\cite{SM,BJL,BBH,BBL}).
There remains some uncertainties
in the prediction due to the quark masses, especially $m_s$,
the QCD scale $\Lambda_{QCD}$, and certain hadronic 
matrix elements. One quoted range is\cite{BJL}:
\begin{equation}
\mbox{Re}\left( \frac{\epsilon^{'}}{\epsilon} \right) 
= (3.6 \pm 3.4) \times 10^{-4}
\label{ep}
\end{equation}
In particular, a vanishing result results from an accidental cancellation
(rather than a symmetry).

The parameter $\epsilon'$ is given by
\begin{equation}
\epsilon' = - 
\frac{ e^{i\left(\pi/2 + \delta_2 - \delta_0\right)} }{\sqrt{2}}
\frac{\mbox{Re}A_2}{\mbox{Re}A_0}
\left[ \frac{\mbox{Im}A_0}{\mbox{Re}A_0} - 
\frac{\mbox{Im}A_2}{\mbox{Re}A_2} \right]
\ ,
\end{equation}
where $A_I$ are the isospin amplitudes in $K\rightarrow\pi\pi$ decays
and $\delta_I$ are the corresponding final state interaction phases.

\begin{figure}[bthp]
\begin{center}
\ \epsfbox{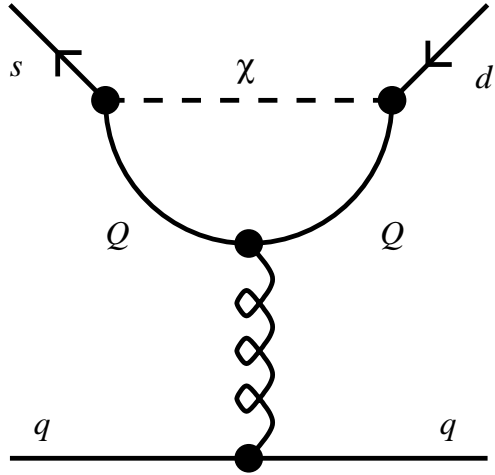}
\end{center}
\caption[]{
Gluon penguin diagram contribution to the imaginary part of the
$K\rightarrow\pi\pi$ decay for the models of soft CP breaking.
}\label{penguin}
\end{figure}

To estimate the contributions to the imaginary part of the
$K\rightarrow\pi\pi$ decay for the models of soft CP breaking,
let us consider the gluon penguin diagram shown in Fig.~\ref{penguin}.
For Type R soft CP breaking model 
the chiralities of $s$ and $d$ quarks in the external lines are
different with those for the $W$-exchange contribution. Then it is
convenient to define the following operators:
\begin{eqnarray}
Q'_3 &=& 4 \left( \bar{s}_R \gamma_\mu d_R \right)
\sum_{q=u,d,s} \left( \bar{q}_R \gamma^\mu q_R \right) \ ,
\nonumber\\
Q'_4 &=& 4 \sum_{q=u,d,s} \left( \bar{s}_R \gamma_\mu q_R \right)
\left( \bar{q}_R \gamma^\mu d_R \right) \ ,
\nonumber\\
Q'_5 &=& 4 \left( \bar{s}_R \gamma_\mu d_R \right)
\sum_{q=u,d,s} \left( \bar{q}_L \gamma^\mu q_L \right) \ ,
\nonumber\\
Q'_6 &=& - 8 \sum_{q=u,d,s} \left( \bar{s}_R q_L \right)
\left( \bar{q}_L d_R \right) \ .
\end{eqnarray}
By using these operators, 
the $\Delta S = 1$ effective Hamiltonian 
for Type R soft CP breaking model 
is given by
\begin{equation}
{\cal H}_{\Delta S=1}^{\rm(new)} = 
\frac{1}{v^2} \overline{C}_{sd}^{(R)}
\sum_{i=3}^6 \nu_i(M_{\rm new}) Q'_i(M_{\rm new})
\ ,
\label{S1}
\end{equation}
where $M_{\rm new}$ is a scale around masses of new particles, and
\begin{equation}
-\frac{1}{3}\nu_3(M_{\rm new}) =
\nu_4(M_{\rm new}) =
-\frac{1}{3}\nu_3(M_{\rm new}) =
\nu_6(M_{\rm new}) =
\frac{\alpha_s(M_{\rm new})}{256\pi} 
\ .
\end{equation}
Here $\overline{C}_{sd}^{(R)}$ is expressed as
\begin{equation}
\overline{C}_{sd}^{(R)} = \frac{v^2}{M_Q^2}
\sum_{I=1}^2 X^I_{sd} \,
\tilde{F} \left( \frac{M_I^2}{M_Q^2} \right) 
\ ,
\end{equation}
where
\begin{equation}
\tilde{F} \left( r_I \right) =
\frac{4}{3(1-r_I)}
\left[
  \frac{7-29r_I+16r_I^2}{6(1-r_I)^2}
  - \frac{r_I(3-2r_I)}{(1-r_I)^3} \ln r_I
\right]
\ .
\end{equation}
The effective Hamiltonian for the Type L soft CP breaking is obtained
by switching the helicities R to L and L to R in the above
expressions, and $\overline{C}_{sd}^{(L)} = \overline{C}_{sd}^{(R)}$.

To obtain the amplitudes for $K\rightarrow \pi\pi$ we need to study
the renormalization group evolution of the coefficients.
This is done by using the method described in, {\it e.g.}
Refs.~\cite{BBH,BBL}.
The resultant coefficients are
$\left(\nu_3(m_c),\nu_4(m_c),\nu_5(m_c),\nu_6(m_c)\right) =
\left( -1.2 , 1.5 , 0.8 , 4.7 \right)\times 10^{-4}$,
where we have taken $M_{\rm new}=M_W$ for simplicity.
As is well known, the gluon penguin diagram gives a contribution to only
isospin zero channel.
We use the values in Ref.~\cite{BBL} for the hadron matrix elements:
$\left( \langle Q'_3(m_c) \rangle_0 , \langle Q'_4(m_c) \rangle_0 ,
\langle Q'_5(m_c) \rangle_0 , \langle Q'_6(m_c) \rangle_0 \right) =
\left( -0.01 , -0.19 , 0.09 , 0.28 \right)$\,$(\mbox{GeV})^3$.
By using the experimental values
$\mbox{Re}A_0=3.33\times10^{-7}$\,GeV and
$\mbox{Re}A_2=1.50\times10^{-8}$\,GeV with
$\left\vert \epsilon_K \right\vert = 2.36 \times 10^{-3}$,
$\mbox{Re} \left(\epsilon'/\epsilon \right)$ 
from the gluon-penguin diagram is given by
\begin{equation}
\left\vert \mbox{Re} \left( \frac{\epsilon'}{\epsilon}
\right) \right\vert
\simeq
\left( 7.7 \times 10^{-2} \right) 
\left\vert \mbox{Im} \overline{C}_{sd}^{(R,L)} \right\vert
\ .
\label{prediction}
\end{equation}
Assuming that the real and imaginary parts of the Yukawa coupling are
comparable, and using the value in Eq.~(\ref{X}) estimated from
$\epsilon_K$, we obtain
\begin{equation}
\left\vert \mbox{Re} \left( \frac{\epsilon'}{\epsilon}
\right) \right\vert
\simeq 2 \times 10^{-5}
\frac{v}{M_Q} \lesssim 2 \times 10^{-5}
\end{equation}
for the models of soft CP breaking.
Note that $\mbox{Re}(\epsilon^\prime/\epsilon)$ can be of order
$10^{-4}$ if we allow that the imaginary part is bigger than the real
part of $X_{sd}^I$, $\mbox{Im}X_{sd}^I \sim 10 \times 
\mbox{Re}X_{sd}^I$.
The prediction in Eq.(\ref{prediction})
is more reliable than the corresponding
prediction in the standard model because there is no expectation
of delicate cancellation between diagrams.

In the Aspon model the dominant contribution is given by the Aspon
gauge boson--heavy quark exchange penguin diagram, and 
$\mbox{Re} \left( \epsilon' / \epsilon \right)$ is estimated 
as~\cite{FH1}
$\mbox{Re} \left( \epsilon' / \epsilon \right) \lesssim 10^{-5}$.

\section{$B^0-\bar{B}^0$ Mixing}
\label{B}

In addition to the $W$-exchange box diagram contribution
the scalar-heavy quark exchange box diagram contribute to 
$B_d$--$\bar{B}_d$ mixing in the models of soft CP violation.
The effective Hamiltonian derived from the
new contribution for the Type R soft CP breaking model
takes the same form as that for $\Delta S=1$ effective Hamiltonian
given in Eq.~(\ref{new H}) with $s$ replaced by $b$, and similarly for
the Type L soft CP breaking model and the Aspon model.
This should be compared with the $W$-top exchange diagram
contribution, which takes the same form as that in Eq.~(\ref{KM H})
with $s$ and $c$ replaced by $b$ and $t$.
Again it may be enough to assume that the two operators with different
chiralities give the same hadron matrix elements.
Then let us compare $C_{bd}$ with $C_{bd}^{\rm(KM)}$.

The experimental value of the top quark mass, $m_t=175$\,GeV, gives
$C_{bd}^{\rm (KM)} \simeq \left(3.46\times 10^{-3} \right)
\left( V_{tb} V_{td} \right)$.
In the models of soft CP breaking 
the quark mixing matrix is real and orthogonal,
and the unitarity triangle is flat.
In the Aspon model the imaginary parts of the mixing matrix arise from the
imaginary parts of the small quantities $x_i$, and the $3\times3$
submatrix is real and orthogonal in good approximation.
So the current experimental value 
$\left\vert \left(V_{ud}^\ast V_{ub}\right) 
/ \left( V_{cd}^\ast V_{cb} \right) \right\vert \simeq 0.35$ leads
$\left\vert \left(V_{td}^\ast V_{tb}\right) 
/ \left( V_{cd}^\ast V_{cb} \right) \right\vert \simeq 0.65$, and
$\left\vert V_{td}^\ast V_{tb} \right\vert \simeq 5.9 \times 10^{-3}$.
This implies $C_{bd}^{\rm (KM)} \simeq 1.2 \times 10^{-7}$.
On the other hand, when we assume that the Yukawa couplings are 
independent of the generation in the models for soft or
spontaneous CP violation considered in this paper,
$C_{bd}$ is roughly of the same order as $C_{sd}$;
$\left\vert C_{bd}^{(R,L,A)} \right\vert \sim
\left\vert C_{sd}^{(R,L,A)} \right\vert \sim 10^{-10}$.
This value is much smaller than $C_{bd}^{\rm (KM)}$, and negligible.
This situation is similar to $\eta=0$ in the standard model,
which is not excluded by the experiment~\cite{FG,BHSW,M}.

In the case of generation-{\it independent} Yukawa couplings, CP
violation in $B_d$--$\bar{B}_d$ mixing is much smaller for the
soft and spontaneous CP breaking models than that for the SM.
On the other hand, we can admit generation-{\it dependent} Yukawa
couplings, and expect that $C_{bd}$ is larger and roughly of the same
order as $C_{bd}^{\rm(KM)}$;
$\left\vert C_{bd}^{(R,L,A)} \right\vert \simeq 10^{-7}$.
For the models of soft CP breaking this corresponds to: 
\begin{equation}
\frac{v}{M_Q} \left\vert X_{bd} \right\vert \simeq 7 \times
10^{-3} \ ,
\label{Xbd}
\end{equation}
where $X_{bd}$ is the average value of $X_{bd}^{I=1}$ and
$X_{bd}^{I=2}$.
[Note that $\mbox{Im} X_{bd} = \mbox{Im} X_{bd}^{I=1} =
- \mbox{Im} X_{bd}^{I=2}$.]
For the Aspon model $\left\vert C_{bd}^{(A)} \right\vert
\simeq 10^{-7}$ leads
\begin{equation}
\frac{v}{\kappa} \left\vert x_1^\ast x_3 \right\vert 
\simeq 2 \times 10^{-4} \ .
\label{x13}
\end{equation}

For the Type R soft breaking model
$\mbox{Im}X_{bd}$ is strongly constrained by $\bar{\Theta}$,
$\left\vert \mbox{Im}X_{bd} \right\vert \lesssim
2 \times 10^{-4}$ (see Eq.~(\ref{thetaR})).
So the above constraint (\ref{Xbd})
for $\left\vert X_{bd} \right\vert$ leads that 
$\left\vert \mbox{Re} X_{bd}
\right\vert$ is much bigger than $\left\vert \mbox{Im}
X_{bd} \right\vert$.  This implies that the CP violation
in $B_d$--$\bar{B}_d$ mixing in the Type R soft breaking
model is much smaller than that in the SM even if we introduce the
generation-dependent Yukawa coupling.
For the Type L soft breaking model and the Aspon model, 
however, the constraint from $\bar{\Theta}$ is not strong,
so that the CP violation in the $B_d$--$\bar{B}_d$ mixing
can be as big as in the SM.

Similarly, for $B_s$--$\bar{B}_s$ mixing, 
we may expect that $C_{bs}$ is as large as $C_{bs}^{\rm (KM)}$.
In such a case,
the CP violation in the $B_s$--$\bar{B}_s$ mixing
for the Type L soft breaking model and the Aspon model
can be as large as in the SM.
On the other hand, due to the constraint 
from $\bar{\Theta}$,
for the Type R soft breaking model it is much smaller
than that in the SM.

\section{Neutral $B$ Decays and CP Asymmetries.}
\label{decays}

The CP violation in 
the neutral $B$ meson decays is expressed by the product of the two
quantities measuring the indirect and direct CP violations,
respectively:
\begin{equation}
\lambda(B_q \rightarrow X) = 
\left( \frac{q}{p} \right)_{B_q} 
\frac{\bar{A}(\bar{B_q} \rightarrow \bar{X})}%
{A(B_q \rightarrow X)}
\ , 
\label{def:lambda}
\end{equation}
where $B_q$ is $B_d$ or $B_s$.
In the SM this quantity measures the angles of the unitarity triangle.
This corresponds to the terms in the requirement that:
\begin{equation}
V_{ub}^{\ast} V_{ud} + 
V_{cb}^{\ast} V_{cd} + 
V_{tb}^{\ast} V_{td} = 0.
\end{equation}
The angles between the 1st \& 2nd, 2nd\& 3rd, and 3rd \& 1st terms are
called $\gamma,
\alpha$, and $\beta$ respectively. The KM model
predicts a sizeable area of the triangle involving,
{\it e.g.} $\sin2\beta > 0.65$\cite{M}.

To study the quantity $\lambda$ in Eq.~(\ref{def:lambda}) in 
the soft and spontaneously broken models,
let us
consider four cases for the coefficients of the 4-fermi
operator as in Eqs.~(\ref{new H}) and (\ref{KM H}). (An alternative
analysis of new physics and the quantity $\lambda$ is in \cite{SW}.)

The first case
corresponds to generation-independent Yukawa couplings. The
other three cases involve generation dependence, in
particular where the third generation couples more strongly
than the second (case 2), the first (case 3) or both (case 4);
these lead, in general, to a deviation from pure superweak
phenomenology. The four cases are explicitly:
\begin{enumerate}
\item
$\left\vert \mbox{Im} C_{bd} \right\vert \sim
\left\vert \mbox{Im} C_{bs} \right\vert \sim
\left\vert \mbox{Im} C_{sd} \right\vert$;

\item
$\left\vert \mbox{Im} C_{bs} \right\vert \sim
\left\vert \mbox{Im} C_{sd} \right\vert$ and
$\left\vert \mbox{Im} C_{bd} \right\vert \sim
\left\vert C_{bd}^{\rm (KM)} \right\vert$;

\item
$\left\vert \mbox{Im} C_{bd} \right\vert \sim
\left\vert \mbox{Im} C_{sd} \right\vert$ and
$\left\vert \mbox{Im} C_{bs} \right\vert \sim
\left\vert C_{bs}^{\rm (KM)} \right\vert$;

\item
$\left\vert \mbox{Im} C_{bd} \right\vert \sim
\left\vert C_{bd}^{\rm(KM)} \right\vert$ and
$\left\vert \mbox{Im} C_{bs} \right\vert \sim
\left\vert C_{bs}^{\rm(KM)} \right\vert$;

\end{enumerate}

The first factor
$(q/p)_{B_q}$ in Eq.~(\ref{def:lambda}) measures the indirect CP 
violation.  In the present models, up to corrections of order
$10^{-2}$, it is given by the quantity with modulus one,
\begin{equation}
\left( \frac{q}{p} \right)_{B_q} \simeq
\frac{C_{bq}^{\rm(KM)} + C_{bq}}%
{\left\vert C_{bq}^{\rm(KM)} + C_{bq}\right\vert}
\label{qpval}
\ .
\end{equation}
Then for $\mbox{Im} C_{bq} \sim \mbox{Im} C_{sd}$ we find
$\mbox{Im}\left( q/p \right)_{B_q} \lesssim 10^{-2}$.
Note that any non-vanishing value for 
$\mbox{Im}\left( q/p \right)_{B_q}$ comes from the approximation
involved in Eq.~(\ref{qpval}).
On the other hand, for $\left\vert \mbox{Im} C_{bq} \right\vert
\sim \left\vert C_{bq}^{\rm(KM)} \right\vert$ as in 2, 3 and 4 (as in
the SM), which is possible 
for the Aspon model and the Type L soft breaking model,
the $\mbox{Im} C_{bq}$ becomes less negligible and so
it is convenient to define
\begin{equation}
\left( \frac{q}{p} \right)_{B_q} \simeq
e^{i 2 \tilde{\beta}_q} \ .
\end{equation}

The second factor
$\left(\bar{A}/A\right)$ in Eq.~(\ref{def:lambda}) measures direct
CP violation in neutral $B$ meson decays.
Neutral $B$ meson decays are described by $\bar{b}\rightarrow q'
\bar{q'} q''$ at the quark level.
In this case the ratio of
$W$-exchange penguin contribution to the tree contribution is
roughly\cite{Neubert}
\begin{equation}
\frac{A_{\rm penguin}^{\rm(KM)}}{A_{\rm tree}} \sim
\left( \mbox{4-10\%} \right) 
\frac{V_{tb}^* V_{tq''}}{V_{q'b}^* V_{q'q''}} \ .
\label{ratio KM}
\end{equation}
In addition, there is a contribution from the scalar-heavy quark
exchange penguin diagram in the soft breaking models, and
a contribution from the Aspon gauge boson-heavy quark
exchange penguin diagram in the Aspon model.
The ratio of the new penguin contribution to 
the $W$-top penguin contribution is given by
\begin{equation}
\frac{A_{\rm penguin}^{\rm(new)}}{A_{\rm penguin}^{\rm(KM)}}
\, \sim \,
\frac{\overline{C}_{bq''}}{\overline{C}_{bq''}^{\rm (KM)}} \ ,
\label{ratio new}
\end{equation}
where $\overline{C}_{bq''}$ and $\overline{C}_{bq''}^{\rm (KM)}$
are analogues of $\overline{C}_{sd}$ in Eq.~(\ref{S1}).
This ratio is estimated by the ratio of the couplings:
\renewcommand{\arraystretch}{1.5}
\begin{equation}
\frac{\overline{C}_{bq''}}{\overline{C}_{bq''}^{\rm (KM)}}
\, \sim \,
\left\{ 
\begin{array}{ll}
\displaystyle
\left( \frac{v}{M_Q} \right)^2 
\frac{X_{bq''}}{V_{tb}^\ast V_{tq''}} \ , \qquad
& \mbox{for the soft breaking models}\ ,
\\
\displaystyle
\left( \frac{v}{\kappa} \right)^2 
\frac{x_3 x_{q''}^\ast}{V_{tb}^\ast V_{tq''}} \ , \qquad
& \mbox{for the Aspon model}\ ,
\end{array}
\right.
\end{equation}
\renewcommand{\arraystretch}{1}
where $x_{d,s} = x_{1,2}$.
For $\mbox{Im} C_{bq''} \sim \mbox{Im} C_{sd}$,
the imaginary part of
this ratio is very small, and the new contribution 
is negligible compared with the KM-penguin
contribution.
When $\left\vert \mbox{Im} C_{bq''} \right\vert
\sim \left\vert C_{bq''}^{\rm (KM)} \right\vert$, 
the imaginary part of this ratio
can be of order one in the Type L soft breaking model,
while it is small, $\lesssim 10^{-1}$, in the Aspon model.
Then if
\begin{equation}
\left\vert \frac{V_{tb} V_{tq''}}{V_{q'b} V_{q'q''}}
\right\vert \leq 1 \ ,
\end{equation}
the tree diagram dominates over penguin 
diagram~\cite{Neubert}, and the direct CP violation
in the $B$ system is small.
This corresponds to the processes $b\rightarrow c\bar{c}s$,
$b\rightarrow c\bar{c}d$ and $b\rightarrow u\bar{u}d$.
On the other hand, if tree diagrams are forbidden, 
the penguin diagram dominates, and
\begin{equation}
\frac{\bar{A}}{A} \sim
\frac{
 \overline{C}_{bq''}^{\rm (KM)} + \overline{C}_{bq''}^\ast
}{
 \overline{C}_{bq''}^{\rm (KM)} + \overline{C}_{bq''}
}
\ .
\end{equation}
This is for $q'=d$ or $s$.
When $\left\vert \mbox{Im} C_{bq''} \right\vert \sim
\left\vert C_{bq''}^{\rm (KM)} \right\vert$
in this case, it is convenient to parameterize
\begin{equation}
\frac{\bar{A}}{A} \simeq e^{i2\tilde{\alpha}_{q''}} \ ,
\end{equation}
where $\tilde{\alpha}_{q''}$ is of order one in the
Type L soft breaking model, $\lesssim 10^{-1}$ in the 
Aspon model, and very small in the Type R soft
breaking model.

\begin{table}[htbp]
\begin{tabular}{ccccccc}
& & (1) & (2) & (3) & (4) & SM \\
\hline
$b \rightarrow c\bar{c}s$ 
 & $B_d \rightarrow \psi K_S$ & 0 & $\sin2\tilde{\beta}_d$
 & 0 & $\sin2\tilde{\beta}_d$ & $-\sin 2 \beta $ \\
& $B_s \rightarrow D^+_s D^-_s$ & 0 & 0 & $\sin2\tilde{\beta}_s$
 & $\sin2\tilde{\beta}_s$ & $-\sin 2 \beta' $ \\
\hline
$b \rightarrow c\bar{c}d$ 
 & $B_d \rightarrow D^+ D^-$ & 0 & $\sin2\tilde{\beta}_d$
 & 0 & $\sin2\tilde{\beta}_d$ & $-\sin 2 \beta $ \\
& $B_s \rightarrow \psi K_S$ & 0 & 0 & $\sin2\tilde{\beta}_s$
 & $\sin2\tilde{\beta}_s$ & $-\sin 2 \beta' $ \\
\hline
$b \rightarrow c\bar{c}s$ 
 & $B_d \rightarrow \pi^+\pi^-$ & 0 & $\sin2\tilde{\beta}_d$
 & 0 & $\sin2\tilde{\beta}_d$ & $\sin 2 \alpha $ \\
& $B_s \rightarrow \rho K_S$ & 0 & 0 & $\sin2\tilde{\beta}_s$
 & $\sin2\tilde{\beta}_s$ & $-\sin 2 (\gamma+\beta') $ \\
\hline
$b \rightarrow s\bar{s}s$ 
 & $B_d \rightarrow \phi K_S$ & 0 & $\sin2\tilde{\beta}_d$
 & $\sin2\tilde{\alpha}_s$ 
 & $\sin2\left(\tilde{\beta}_d+\tilde{\alpha}_s\right)$ 
 & $-\sin 2 (\beta-\beta') $ \\
& $B_s \rightarrow \eta' \eta'$ & 0 & 0 
 & $\sin2\left(\tilde{\beta}_s+\tilde{\alpha}_s\right)$
 & $\sin2\left(\tilde{\beta}_s+\tilde{\alpha}_s\right)$
 & 0 \\
\hline
$b \rightarrow s\bar{s}d$ 
 & $B_d \rightarrow K_S K_S$ & 0 
 & $\sin2\left(\tilde{\beta}_d+\tilde{\alpha}_d\right)$
 & 0 
 & $\sin2\left(\tilde{\beta}_d+\tilde{\alpha}_d\right)$ 
 & 0 \\
& $B_s \rightarrow \phi K_S$ & 0 & $\sin2\tilde{\alpha}_d$
 & $\sin2\tilde{\beta}_s$
 & $\sin2\left(\tilde{\beta}_s+\tilde{\alpha}_d\right)$
 & $\sin 2 (\beta-\beta')$ \\
\end{tabular}
\caption[]{
Values of $\mbox{Im}\lambda\left(B_q\rightarrow X\right)$
for the examples of the neutral $B$ meson decay modes.
(1)--(4) correspond to four cases discussed in text.
A zero indicates that the value is small, 
$\lesssim{\cal O}(10^{-2})$.
The column indicated by ``SM'' shows the predictions in 
the SM~\cite{Neubert}.
}\label{tab:1}
\end{table}

In Table~\ref{tab:1} we show examples of neutral $B$ 
meson decay modes with values of $\mbox{Im}\lambda
\left(B_q \rightarrow X\right)$ for the four cases
discussed above.
One can read from Table~\ref{tab:1}
specific features of the present models.
For example: if CP assymetry in $B_d \rightarrow K_S K_S$ were large,
then it indicates a clear deviation from the Standard
Model, and those for tree dominant decay modes are the same;
$\mbox{Im}\lambda\left(B_d \rightarrow \psi K_S\right)
\simeq \mbox{Im}\lambda\left(B_d \rightarrow D^+ D^-\right)
\simeq \mbox{Im}\lambda\left(B_d \rightarrow \pi^+ \pi^-\right)$.
On the other hand, if it were small, all CP violations
in $B_d$ decays are small.

If we focus just on the ``gold-plated'' decay mode
$B \rightarrow \psi K_S$ (top row of Table~\ref{tab:1})
- where the SM predicts an unmistakable large CP asymmetry -
then in the Type R soft breaking model one must have
condition (1) and hence a very small $\beta$ ($\beta < 10^{-2}$); 
in the Type L soft breaking model or the Aspon model one
{\it can} admit conditions (2) and (4) and hence large 
effective $\beta$.
However, if we impose that the Yukawa couplings are 
generation-independent, all except the SM
predict a CP asymmetry in this mode too
small to be detected.

\section{Compatibility with Upper Bound on $\bar{\Theta}$;
{\it Lower} Bounds on Electric Dipole Moments.}
\label{theta}

It is interesting to estimate the {\it lower} bound
on $\bar{\Theta}$ and hence on the neutron electric dipole
moment $d_n$, for the different models.

First recall that in the standard model where the strong CP
problem is unresolved - and requiring an
additional mechanism such
as the Peccei-Quinn symmetry~\cite{PQ} or a massless up quark 
(see, {\it e.g.} Ref.~\cite{P}) -
there is no such lower limit
because there is no reason to make $\bar{\Theta}$
small. If one simply puts the bare $\bar{\Theta}$
equal to zero (without motivation) then it has been
pointed in Ellis and Gaillard\cite{EG} that
there is a finite correction at two loops of $\sim 10^{-16}$
and an infinite renormalization at seven loops
which is even smaller, $\sim 10^{-32}$ if one arbitrarily
puts in a cut-off equal to the Planck mass. But these are not really
predictions for 
a lower bound because there is fine-tuning unless there is an additional 
mechanism.

The value of $\bar{\Theta}$ is strongly constrained by
the experiment of the neutron electric dipole moment;
$\bar{\Theta}\leq 10^{-10}$.
In the models considered in this paper the determinants of
mass matrices of quarks are real, and
the resultant $\bar{\Theta}$ is zero at tree level.
However, it is generated at some loop level through
corrections to the mass matrix;
$\bar{\Theta} = 
\mbox{Im} \left\{
\mbox{tr}\left[ M^{-1} \delta M \right] \right\}$,
where $M$ is the tree level mass matrix.

In the Aspon model a contribution to $\bar{\Theta}$
appears at one-loop level due to the mixing between the
heavy scalar $\chi_I$ and the ordinary Higgs boson $H$
given in Eq.~(\ref{L chi H})~\cite{FN}.
This contribution is estimated as~\cite{FG}
\begin{equation}
\bar{\Theta} = \frac{\lambda x^2}{16\pi^2} \ ,
\end{equation}
where $\lambda$ is an average value of $\lambda_{IJ}$
in Eq.~(\ref{L chi H}) and $x$ is an average value of 
$\left\vert x_i \right\vert$ in Eq.~(\ref{def x}).
{}From a one-loop correction from the 
quark box diagram 
a lowest value of $\lambda$ and hence
$\bar{\Theta}$ are estimated as~\cite{FH}
\begin{equation}
\lambda \gtrsim \frac{x^2}{16\pi^2}\ ,
\qquad
\bar{\Theta} \gtrsim \frac{x^4}{\left(16\pi^2\right)^2} \ .
\label{ThetaA}
\end{equation}
This by using the upper bound of $\bar{\Theta}$
implies $x^2 \lesssim 10^{-3}$.
When the Yukawa couplings are generation-independent, 
we obtain $\kappa \lesssim 3 \times 10^4$\,GeV by combining
this with the constraint (\ref{eps}) from $\epsilon_K$.
The assumption $\kappa > v$ gives the lower bound $x^2 \gtrsim
10^{-5}$, which leads to $\bar{\Theta} \gtrsim 4 \times 10^{-15}$, and
hence $d_n \gtrsim 4 \times 10^{-30}$e.cm.
As discussed in Section(\ref{B}) one can admit 
$\left\vert C_{bd}^{(A)} \right\vert$ is as large as 
$\left\vert C_{bd}^{\rm(KM)}\right\vert$ by using the
generation-dependent Yukawa couplings.  In such a case the combination
of Eqs.~(\ref{ThetaA}) with
the constraint (\ref{x13}) from $B_d$--$\bar{B}_d$
mixing, we obtain $\kappa \lesssim 10^3$\,GeV (rather than $\kappa
\lesssim 3 \times 10^4$ GeV).
Equation~(\ref{x13}) with the assumption $\kappa > v$ gives
the lower bound $\left\vert x_1^\ast x_3 \right \vert \gtrsim 2 \times
10^{-4}$, which combined with Eq.~(\ref{ThetaA}) leads $\bar{\Theta}
\gtrsim 2 \times 10^{-12}$, and hence $d_n \gtrsim 2 \times
10^{-27}$e.cm.

\begin{figure}[bthp]
\begin{center}
\ \epsfbox{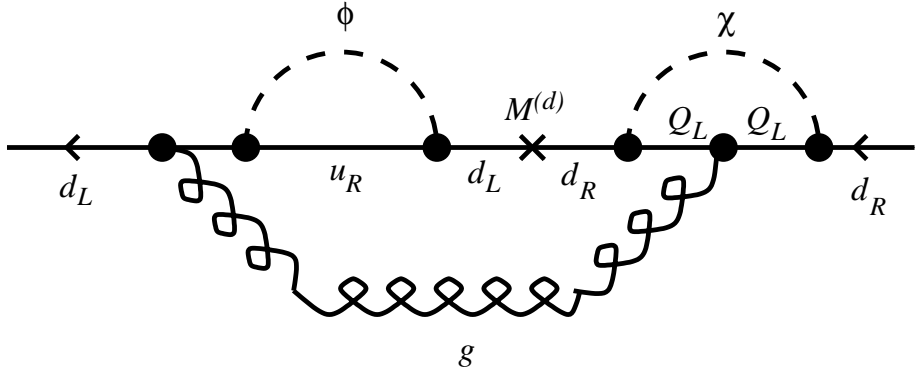}
\end{center}
\caption[]{%
Three loop diagram which gives a correction 
to the imaginary part of the $d$ mass matrix
in the model for Type R soft CP breaking.
}\label{fig:theta R d}
\end{figure}

In the model for Type R soft CP breaking
the corrections to the imaginary part of the mass matrix of the down
sector first arise at two-loop level, as pointed out in ~\cite{BCK}.
We estimate this as $\bar{\Theta} \simeq \lambda f^2 / 
\left(16\pi^2\right)^2$, where $\lambda$ is an average value of
$\lambda_{IJ}$ in Eq.~(\ref{L chi H}) and $f$ is an average value of
the Yukawa couplings.  Different from the case of the Aspon model 
(where the top quark contributes), a
lowest value of $\lambda$ is here estimated from a one-loop correction from
the box diagrams of {\it down-type} quarks.
The resultant lowest value of $\bar{\Theta}$ is estimated as
$\bar{\Theta} \gtrsim \left( f^4/(16\pi^2)^3 \right) \cdot
\left( m_b / v \right)^2 $, which leads only to $f^2\lesssim1$.
This constraint is not strong.  However, a stronger constraint comes
from the three-loop diagram shown in Fig.~\ref{fig:theta R d}.
The correction from the diagram to the imaginary part of the mass
matrix of down sector is estimated as
\begin{equation}
\delta {\cal M}_d \sim \frac{\alpha_s}{4\pi}
\frac{1}{16\pi^2 v^2} \frac{1}{16\pi^2}
V^T \left( {\cal M}_u \right)^2 V {\cal M}_d X^I
\ ,
\end{equation}
where $\alpha_s \simeq 0.12$ is the QCD coupling
and $V$ is a real orthogonal KM matrix.
The contribution to 
$\bar{\Theta}$ is calculated by multiplying the above
$\delta {\cal M}_d$ by $\left( {\cal M}_d \right)^{-1}$
and taking trace.  Since ${\cal M}_d$ is included between
two hermitian matrices in $\delta {\cal M}_d$, enhancement
factors arises from $\left( {\cal M}_d \right)^{-1}$.
The resultant correction to $\bar{\Theta}$ is given by
\begin{eqnarray}
\bar{\Theta}(\mbox{down}) &\sim&
\frac{\alpha_s}{4\pi} 
\frac{m_t^2}{ \left(16\pi^2\right)^2 v^2 }
\left[
  \frac{m_s}{m_d} V_{ts} V_{td}\, \mbox{Im}X^I_{12}
  + \frac{m_b}{m_d} V_{tb} V_{td}\, \mbox{Im}X^I_{13}
  + \frac{m_b}{m_s} V_{tb} V_{ts}\, \mbox{Im}X^I_{23}
\right] \ .
\end{eqnarray}
By using $(m_d,\,m_s,\,m_b) \simeq
(8,\,150,\,4800)$\,MeV and $(V_{td},\,V_{ts},\,V_{tb})
\simeq (5.9\times10^{-3},\,4.3\times10^{-2},\,1)$,
the above expression becomes
\begin{equation}
\bar{\Theta}(\mbox{down}) \sim
\left[
  9.0\times 10^{-3} \mbox{Im} X^I_{sd}
  + 6.7 \mbox{Im} X^I_{bd} + 2.6 \mbox{Im} X^I_{bs}
\right] \times 10^{-7}
\ .
\end{equation}
Then the constraint $\bar{\Theta} \lesssim 10^{-10}$ gives
\begin{eqnarray}
&& \left\vert \mbox{Im} X^I_{sd} \right\vert
\lesssim 0.1 \ ,\nonumber\\
&& \left\vert \mbox{Im} X^I_{bd} \right\vert
\lesssim 2\times10^{-4} \ ,
 \nonumber\\
&& \left\vert \mbox{Im} X^I_{bs} \right\vert
\lesssim 4\times10^{-4} \ .
\label{thetaR}
\end{eqnarray}
When we demand $\left\vert C_{bd}^{(R)}\right\vert \simeq 
\left\vert C_{bd}^{\rm(KM)} \right\vert$ consistently with the above
upper bound,
we need to require 
$\left\vert \mbox{Re}X^I_{bd} \right\vert \gg 
\left\vert \mbox{Im}X^I_{bd} \right\vert$.
Then the bounds for $M_Q$ and $\bar{\Theta}$ are same as those for the
generation-independent Yukawa couplings.
The combination of the upper bound (\ref{thetaR}) with the constraint
obtained from $\epsilon_K$ (Eq.~(\ref{X})) gives the upper bound for
$M_Q$: $M_Q^{(R)} \lesssim 8 \times 10^4$\,GeV.
The lower bound for $\bar{\Theta}$ may be obtained from the lower
bound for $\bar{X}_{sd}$ ($\bar{X}_{sd} \gtrsim 3 \times 10^{-4}$)
derived from $\epsilon_K$.  The result is
$\bar{\Theta} \gtrsim 3 \times 10^{-13}$, and hence
$d_n \gtrsim 3 \times 10^{-28}$.

In the model of Type L soft CP breaking, a contribution to 
$\delta {\cal M}$ arises at two-loop level\footnote%
{This two-loop contribution is due to Sheldon Glashow.}
from the diagram similar to
the one for the Type R soft breaking model,
while the three-loop diagram similar to the one in 
Fig.~\ref{fig:theta R d} does not contribute to
$\bar{\Theta}$.  So the dominant contribution to $\bar{\Theta}$ is
estimated as $\bar{\Theta} \simeq \lambda f^2/ (16\pi^2)^2$.
Similarly to the Aspon model, a lowest value of $\lambda$ is estimated
from a one-loop correction from the box diagram of {\it both} up-type
{\it and} down-type quarks.  The resultant value of $\bar{\Theta}$ is
thus estimated as $\bar{\Theta} \gtrsim f^4/(16\pi^2)^3$, which leads
to $f^2\lesssim 0.02$.
For the case of generation-independent Yukawa couplings the
combination of this upper bound with the constraint from $\epsilon_K$
gives an upper bound for $M_Q$: $M_Q \lesssim 2 \times 10^5$\,GeV.
The bound $\bar{X}_{sd} \gtrsim 3 \times 10^{-4}$ leads to
$\bar{\Theta} \gtrsim 2 \times 10^{-14}$, and hence $d_n \gtrsim 2
\times 10^{-29}$e.cm.
On the other hand, when we require $\left\vert C_{bd}^{(L)}
\right\vert \simeq \left\vert C_{bd}^{\rm(KM)} \right\vert$, the
combination of $f^2 \lesssim 0.02$ with the constraint
obtained from $B_d$--$\bar{B}_d$ mixing gives an
upper bound for $M_Q$: $M_Q \lesssim 7\times 10^2$\,GeV.
A lower bound for $\bar{X}_{bd}$ can be derived
from the $B_d$--$\bar{B}_d$ mass
difference (Section(\ref{B})): 
$\bar{X}_{bd} \gtrsim 7 \times 10^{-3}$, which leads to $f^2\gtrsim
7\times 10^{-3}$.  {}From this lower bound
the lower bound for $\bar{\Theta}$
is estimated as $\bar{\Theta} \gtrsim 10^{-11}$, and hence
$d_n \gtrsim 10^{-26}$e.cm., quite close to the experimental limit.

\section{Summary of Predictions.}
\label{summary}

The predictions of the different models we have
studied are collected together\footnote{%
In Table~\ref{tab:2} the aspon model is ``L-type spontaneous'' meaning
that light left-handed quarks couple to the new quarks. If
we replace this by an aspon model with q= right-handed 
down-type quarks, the lower limits on $\lambda$
and $\bar{\Theta}$ in Eq.(\ref{ThetaA})
are each reduced by a factor $(m_b/v)^2 \sim 10^{-3}$.}
are in Table~\ref{tab:2}.

\begin{table}[ht]
\begin{tabular}{||c||c|c|c|c||}
 &  KM  &  R-type soft  & L-type soft & Aspon \\ 
\hline
\hline
$\left(\frac{\epsilon^{'}}{\epsilon}\right)$ & few $10^{-4}$ &
? & ? & $<10^{-5}$ \\
$\bigtriangleup$ & big & flat & ? & ? \\
$\bar{\Theta}$ & axion? & $>10^{-13}$ ($10^{-13}$)
  & $>10^{-11}$ ($10^{-14}$) & $>10^{-12}$ ($10^{-15}$)\\
$d_n$ & $10^{-32}$ & $>10^{-28}$ ($10^{-28}$)
  & $>10^{-26}$ ($10^{-29}$) & $>10^{-27}$ ($10^{-30}$) \\
\end{tabular}
\caption[]{
Summary of results for the three CP violation models
compared to the SM. 
$\bigtriangleup$ denotes the unitarity triangle determined from
neutral $B$ meson decays.
A query ? denotes not {\it necessarily} pure superweak (essentially
zero $\epsilon'/\epsilon$ and a flat $\bigtriangleup$),
but becomes so if the Yukawa couplings are generation-independent.  In
that case, the first two rows in the last three columns become
indistinguishable.
Values in parentheses denote weaker bounds for the case of
generation-independent Yukawa couplings, to be 
compared to generation-dependent ones.}
\label{tab:2}
\end{table}

{}From this Table we see that the predictions for the different
models are very divergent and therefore when the quantity
$(\epsilon^{'}/\epsilon)$ is measured with an accuracy
of $10^{-4}$, and the CP asymmetry in $B\rightarrow \psi K_S$
is measured to determine whether or not $\sin 2\beta > 10^{-2}$
we will be able to exclude models. As mentioned in the Introduction we
expect that both of these measurements will be completed
within perhaps 2 or 3 years.

\bigskip

\section*{Acknowledgement}

This work was supported in part by the
US Department of energy under Grant No. 
DE-FG05-85ER-40219.

\end{document}